\documentclass[12pt]{article}

\usepackage{newtxtext,newtxmath}
\usepackage{graphicx}

\usepackage[letterpaper,margin=1in]{geometry}

\renewenvironment{abstract}
	{\quotation}
	{\endquotation}

\date{}

\makeatletter
\renewcommand{\fnum@figure}{\textbf{Figure \thefigure}}
\renewcommand{\fnum@table}{\textbf{Table \thetable}}
\makeatother

\usepackage{scicite}

\usepackage{url}

\def\scititle{
	Diffusional earthquakes and their slip-distance scaling
}
\title{\bfseries \boldmath \scititle}

\author{
	Dye~SK~Sato$^{1\ast}$,
	Keisuke~Yoshida$^{2}$\and
	\small$^{1}$
    Research Institute for Marine Geodynamics, Japan Agency for Marine-Earth Science and Technology, 
    \\\small 
    Yokohama 236‑0001, Japan. \and
	\small$^{2}$
    Research Center for Prediction of Earthquakes and Volcanic Eruptions, Graduate School of Science, 
    \\\small 
    Tohoku University, Sendai 980-0845, Japan.\and
	\small$^\ast$Corresponding author. Email: daisukes@jamstec.go.jp%\and
}

\begin{document} 

% Insert the title and author list
\maketitle

% Abstract, in bold
% There are strict length limits, and not all formats have abstracts.
% Consult the journal instructions to authors for details.
% Do not cite any references in the abstract.
\begin{abstract} \bfseries \boldmath
% Start with one or two sentences of background
The final size of an earthquake typically cannot be predicted from its ongoing seismic radiation. Expanding observations reveal distinct exceptions, such as slow earthquakes, injection-induced seismicity, and earthquake swarms, in which fault slip has an upper bound. A common thread among these anomalies is the diffusive migration of their active areas. Here, we report a unified scaling relation for these diffusional earthquakes. By tracking prolonged earthquake swarms in Northeast Japan, we constrained the time evolution of their active seismicity areas and cumulative seismic moments. Their moment-duration trajectories coincide with the final states documented for global swarms and induced seismicity across various scales. When plotted as seismic moment versus seismicity area, their trajectories collapse onto those of slow earthquakes, uniformly explained by a diffusional constant-slip model. This constant-slip scaling carves out a unique class of diffusional earthquakes, where the final available seismic energy is predetermined by slip distance.
%The constant-slip scaling of diffusional earthquakes and the constant-stress-drop scaling of ordinary earthquakes mark a bimodal predictability of seismic energy.
%The final size of an earthquake typically cannot be predicted from its ongoing seismic radiation. Expanding observations reveal distinct exceptions: slow earthquakes, injection-induced seismicity, and earthquake swarms, where the fault slip is bounded. A common thread among these anomalies is the diffusive nature of the maximum migration speeds of their active areas. Here, we report a unified scaling relation for the slip distance of these diffusional earthquakes. By tracking prolonged earthquake swarms in Northeast Japan, we constrained the time evolution of the active seismicity areas and the cumulative seismic moment releases. Their temporal trajectories coincide with the distribution of total moment release versus duration for documented swarms and injection-induced seismicity across various scales. When plotted as seismic moment versus seismicity area, the trajectories of swarms and injection-induced seismicity collapse onto those of slow earthquakes, uniformly explained by a diffusional constant-slip model. The scaling of diffusional earthquakes indicative of a constant slip contrasts with that of ordinary earthquakes based on a constant stress drop, revealing two distinct categories of seismogenesis, namely slip-distance-predictable and stress-drop-predictable earthquakes.
\end{abstract}

% The first paragraph of any Science paper does NOT have a heading
% Nor is it indented
\noindent
The beginning of a large earthquake is so similar to a small one that it precludes the final-size prediction from seismic waveforms, which is the unsettled, similarity hypothesis of ordinary earthquakes. When averaged over individual events, the stress on the source fault drops by roughly 3 MPa, and the rupture propagates almost at 3 km/sec~\cite{kanamori1975theoretical,geller1976scaling}. 
While their typical behavior is well described by a single circular crack~\cite{sato1973body,madariaga1976dynamics}, the intermittent high-frequency excitation observed in large earthquakes indicates that an earthquake is a collective rupture comprising multiple events~\cite{boatwright1982dynamic,boore1983stochastic}. Multi-scale events cascade, yet the stress drop $\Delta \tau$ and the rupture speed $v_{\rm r}$ converge to intrinsic values, a typical process of ordinary earthquake rupture~\cite{ellsworth1995seismic,ide2005earthquakes}. For a circular crack, the relationship between the rupture radius $R$ and the moment release $M_0$ is given by \begin{equation}
    M_0=\frac{16}7 \Delta \tau  R^3,
\end{equation}
and for an ordinary earthquake, with respect to duration $t$,
\begin{equation}
    R= v_{\rm r} t.
\end{equation}
Substituting the typical values of $\Delta \tau$ and $v_{\rm r}$ for ordinary earthquakes yields the third time derivative of the seismic moment common to the growth processes of small and large earthquakes~\cite{uchide2010scaling,meier2016evidence}:
\begin{equation}
  {\rm d}^3 M_0/{\rm d} t^3 \simeq 10^{18}\, [{\rm N}\cdot{\rm m}/{\rm sec}^3] \,\,\,\,\,\,\,\, ({\rm Ordinary\,EQs}).
\end{equation}

Slow earthquakes violate the constancy of $\Delta \tau$ and $v_{\rm r}$ that underlies the similarity hypothesis of ordinary earthquakes. Although slow slip events (SSEs) remain debated~\cite{michel2019similar}, many slow earthquakes, as exemplified by Very Low Frequency Earthquakes (VLFEs) and Secondary Slip Fronts (SSFs), can be described as collective ruptures, much like ordinary earthquakes~\cite{shelly2007non,bletery2017characteristics}. However, the moment release rate of slow earthquakes is bounded, remarkably ``slow'' in contrast to ordinary earthquakes that rupture ``fast''~\cite{ide2007scaling,ide2023slow}:
\begin{equation}
    {\rm d}M_0/{\rm d}t < 10^{13}\, [{\rm N}\cdot{\rm m}/{\rm sec}] \,\,\,\,\,\,\,\, ({\rm Slow\,EQs}).
\end{equation}
Indeed, their source areas do not expand at speeds comparable to the S-wave velocity. Slow earthquakes formed by collective ruptures often migrate, with their propagation speed decreasing with increasing scale, which is apparently diffusive~\cite{ide2023slow,ide2010striations},
\begin{equation}
    R < \sqrt{Dt}\,\,\,\,\,\,\,\, ({\rm Slow\,EQs}). 
\end{equation}
Here, $D$ denotes the diffusion coefficient, which is approximately $10^4$ m${}^2$/sec for slow earthquakes~\cite{ide2010striations,nakano2018shallow}.

In understanding the distinct regularities of fast and slow earthquakes, earthquake swarms may be instrumental as the seismicity that exhibits intermediate characteristics. Earthquake swarms and injection-induced seismicity, both thought to belong to the broader category of fluid-related seismicity~\cite{wei20152012,keranen2018induced}, typically comprise individual rupture events that are ballistic ($R \propto t$, eq. 2) like ordinary earthquakes, whereas their source areas migrate diffusively like certain sorts of slow earthquakes. Namely, when the seismicity zone is defined by the migration front rather than by individual rupture zones, its area expands diffusively~\cite{shapiro1997estimating}:
\begin{equation}
    R\simeq \sqrt{Dt}\,\,\,\,\,\,\,\, ({\rm EQ\,Swarms}). 
\end{equation}
Furthermore, for both earthquake swarms and injection-induced seismicity, the total moment release $M_0$, which is the cumulative sum of individual seismic moments, is proportional to the duration, similar to the upper bound scaling of slow earthquakes~\cite{shapiro2011magnitudes,kwiatek2019controlling,danre2024parallel}. However, the stress drop of individual events within earthquake swarms is comparable to that of ordinary earthquakes, being larger than that of slow earthquakes~\cite{yoshida2017temporal,yoshida2018hypocenter}. The diffusion coefficient $D$ for both earthquake swarms and injection-induced seismicity falls within the order of 0.01--1 m$^2$/sec, which is orders of magnitude smaller than that of slow earthquakes, with substantial variance~\cite{shapiro1997estimating,amezawa2021migration}.

Considering the diffusive nature of their active area envelopes, slow earthquakes may be termed diffusional earthquakes~\cite{ide2007scaling,ide2010striations}, and fluid-related seismicity equally qualifies as such if viewed as a collective rupture rather than mere seismicity. As diffusional collective ruptures, these ``earthquakes'' exhibit parallel trends in seismic moment and duration~\cite{danre2024parallel}. The only significant difference is their diffusion rates. Extending this similarity, we report the following unified scaling relations bridging slow earthquakes and fluid-related seismicity as two similar diffusional earthquakes. (I) While the average rates of moment $M_0/t$ and area $S/t$ (where $S \propto R^2$) differ significantly between the two phenomena, the upper bound $m_*$ of $M_0/S$ is common to diffusional earthquakes:
\begin{equation}
M_0/S < m_* =10^{8.5} \,[{\rm N}\cdot{\rm m}]\,\,\,\,\,\,\,\, ({\rm Diffusional\,EQs}).
\end{equation}
This implies that the characteristic of slow earthquakes, wherein the upper bound of the average slip distance $\bar{\delta}$ calculated as the ratio of $M_0/S$ to the crustal rigidity $\mu=40$ GPa is approximately 1 cm~\cite{ide2007scaling}, directly applies to earthquake swarms and injection-induced seismicity. 
(II) Empirically, the relationship between $M_0$ and $S$ for these diffusional earthquakes is well described by a single scaling function $f$ with a kink at a characteristic area $S_* \sim 10$ [km$^2$]:
\begin{equation}
    M_0= f(S)\simeq m_* S \min(S/S_*, 1) \,\,\,\,\,\,\,\, ({\rm Diffusional\,EQs}).
\end{equation}
Here, $f(S)$ is a saturating function where $M_0/S$ plateaus beyond the characteristic area of 10 km$^2$. This formulates what it means that diffusional earthquakes differ solely in their diffusion rates: from a snapshot of their active areas, slow earthquakes and earthquake swarms are indistinguishable. For ordinary earthquakes, the governing macrophysics that yields the similarity function relating $M_0$ and $S$ was the circular crack model~\cite{sato1973body,madariaga1976dynamics}. The macrophysical model that provides the scaling function for diffusional earthquakes above is the Brownian rupture model~\cite{ide2008brownian}.

Figure 1 illustrates the spatiotemporal evolution of the earthquake swarm at the Yamagata-Fukushima border, triggered by the broad stress field changes following the 2011 $M_w$ 9.0 Tohoku-Oki earthquake. Over an approximately two-year period from March 18, 2011, to May 19, 2013, more than 28,000 earthquakes occurred, releasing a cumulative seismic moment equivalent to an $M_w$ 5.1 event. The relocated hypocenters delineate a distinct planar distribution (Fig. 1a, b), and their migration broadly follows a diffusion coefficient of nearly 1 m${}^2$/sec, which is typical of fluid-related seismicity. Furthermore, the time histories of the seismicity rate, moment release rate, and average moment per event exhibit an exceptionally gradual decay proportional to a negative power of time (-0.2, shown in Fig. S1), reminiscent of injection-induced seismicity during a post-injection phase~\cite{vidale2006survey,langenbruch2010decay}. These observations strongly support the view that, despite its massive scale, this sequence represents a typical earthquake swarm: an accumulation of fault ruptures driven by fluid diffusion intruding into pre-existing weak planes, where individual events have a stress drop on the order of 3 MPa~\cite{yoshida2017temporal,yoshida2018hypocenter}.

The rupture process of this sequence is not a continuous expansion from a single point, but primarily consists of six sub-clusters expanding from their respective initiation points (Fig. 1a, b, colored points; see Methods and Figs. S2--S5 for the clustering procedure). %Those subclusters were objectively identified via density-based clustering (see Methods). 
After filtering out geometrical outliers, the major axis $X_1$ and minor axis $X_2$ of the active region, extracted via principal component analysis, respectively correspond to a strike of N30$^{\circ}$E and the up-dip direction (azimuth S60$^{\circ}$E) of a fault plane dipping 15$^{\circ}$. The spatial extent within this $X_1$-$X_2$ plane is approximately two to four times larger than its thickness in the normal direction $X_3$ (Fig. 1b), indicating that the macroscopic activity diffuses anisotropically along an effective single plane striking N30$^{\circ}$E and dipping 15$^{\circ}$ westward.

Both envelope anisotropy and the clustered nature of the seismicity affect area estimation (Fig. 1c), requiring careful geometric treatment to heighten the scaling accuracy. 
Diffusional anisotropy creates a discrepancy between a circular evaluation ($\pi|X|^2$) and an elliptical one ($\pi|X_1X_2|$), as demonstrated in the 95\%-event envelopes~\cite{amezawa2021migration} of them (Circular and Elliptic, respectively). 
Although these conventional indicators depend on the coordinate origin, this dependence is negligible here because the area of a single convex hull enclosing the entire system (convex polygon; purple dashes in Fig. 1a), which is independent of the origin, tracks the elliptical envelope area. 
However, a single convex hull overestimates the seismicity area due to inactive regions between clusters, as evidenced by the combined area of the primary six sub-clusters (multiple convex polygons) being substantially smaller than the single-hull evaluation. Accordingly, we adopt a concave hull that optimally conforms to the irregular periphery to evaluate the seismicity area of the swarm (Methods). Its temporal evolution (concave polygon; green solid lines in Fig. 1a) is consistent with the sum of the convex hull areas for the six sub-clusters, confirming that this area metric captures the clustered activities.

Figure 2 shows the spatiotemporal evolution of seismic moment release for this earthquake swarm in comparison to those of global slow earthquakes, natural swarms, and injection-induced seismicity ~\cite{michel2019similar,bletery2017characteristics,danre2022prevalence,supino2020self,thomas2016constraints,ide2018seismic}. In the time domain, cumulative moment versus duration, the moment release rate of the Yamagata-Fukushima swarm (diamond in Fig. 2a, calculated from the JMA magnitude, detailed in Methods ~\cite{edwards2009comparative}) is notably small compared to the trend of slow earthquakes (Fig. 2a, circles). 
Furthermore, a similar analysis of the concurrent Sendai-Okura swarm, whose diffusion coefficient is roughly one order of magnitude lower than that of this sequence~\cite{yoshida2018sendai}, reveals a moment rate that is actually more than one order of magnitude lower (up-pointing triangles), highlighting the strong variability of swarm kinematics in the time domain. These time-evolving trajectories are consistent with the final states of other global earthquake swarms and injection-induced seismicity (down-pointing triangles and squares, respectively). Although the scatter is remarkably large, their moment release rates are generally lower than those of slow earthquakes.

When evaluated against the seismicity area, the cumulative moment trajectories collapse onto a single scaling band (Fig. 2b). 
%Here, the plotted areas for previously reported earthquakes were evaluated using circular or elliptical envelopes (see Methods). 
This collapse arises because the ratio of moment rates between slow earthquakes and swarms is nearly equal to the ratio of their diffusion coefficients, implying an identical upper bound in the spatial dimension. The scatter in diffusion coefficients typical of fluid-related seismicity also cancels out. 
For the Yamagata-Fukushima sequence, after passing the initial few tens of data points, where the seismicity area is ill-defined, its trajectory follows that of slow earthquakes from VLFEs to SSFs. Combining the upper-bound scaling of slow earthquakes (eq. 3) with the diffusive spatial propagation $S < \pi D t$ (from eq. 5) yields an areal moment density of $10^{8.5}$ N/m (eq. 7). Assuming a typical crustal rigidity $\mu=40$ GPa, this corresponds to an average fault slip of 1 cm:
\begin{equation}
    \bar\delta :=M_0/(\mu S)\lesssim 1\,{\rm cm}.
\end{equation}
Like the 3 MPa stress-drop scaling for ordinary earthquakes, the average slip of swarms, injection-induced seismicity, and slow earthquakes exhibits an order-of-magnitude scatter but rarely exceeds 1 cm, demonstrating their similarity in the spatial domain.

The constant-slip behavior inferred from the temporal distribution seems to hold true in the spatial domain as well (Fig. 3a). We modeled the spatial distribution of the cumulative slip distance by assuming a circular crack with a 3 MPa stress drop for individual events (see Methods). 
In this first-order model, while the rupture areas of relatively large events of magnitudes 3--4 exhibit features typical of circular cracks with a constant stress drop, assembled smaller events result in a spatially uniform cumulative slip distribution. Within the slip zone, the average cumulative slip distance is on the order of centimeters (1–2 cm, Fig. S6), consistent with the upper bound of the mean slip distance derived above.

In the scaling relation for diffusional earthquakes demonstrated here, the common metric is not the stress drop but the slip distance. Ordinary earthquakes scale with a constant average stress drop, meaning the rate of the mean moment density ${\rm d}(M_0/S)/{\rm d}t$ is constant. A standard circular crack with a crustal rigidity of $\mu=40$ GPa (eq. 1) yields a rate of average slip distance $\bar{\delta}$ on the order of ten centimeters per second (${\rm d}\bar{\delta}/{\rm d}t = \mu^{-1}{\rm d}(M_0/S)/{\rm d}t \simeq 0.2$ m/sec).
The upper-bound scaling for diffusional earthquakes instead sets a threshold for the mean moment density $M_0/S$, which implies a maximum mean slip distance on the order of centimeters for circular ruptures ($\bar{\delta} \lesssim 1$ cm).
The consequent off-fault strain release $\Delta\tau/\mu$ is nearly constant at $10^{-4}$ for ordinary earthquakes; it is proportional to $\bar{\delta}/R$ in diffusional earthquakes, meaning that larger-scale diffusional events result in lower local strain release. 
The strain release bounds scaled energy, radiated energy $E_{\rm R}$ per $M_0$, $E_{\rm R}/M_0 \lesssim \Delta\tau/2\mu$~\cite{kanamori1975theoretical}. Replacing moment with area yields $E_R/S \lesssim \bar{\delta}\Delta\tau/2$, indicating that $\bar \delta\lesssim 1$ cm caps areal radiated-energy density $E_R/S$, restricting equivalent ordinary earthquakes to $M_w$ 4–5.
%Ordinary earthquakes are ``stress-drop predictable'' with a constant off-fault strain release ($\Delta\tau/\mu$), whereas diffusional earthquakes are ``slip-distance predictable,'' governed by a constant on-fault strain release $\bar{\delta}$. 

The concept of a diffusional rupture characterized by a constant slip, anticipated for slow earthquakes~\cite{ide2008brownian,ide2019two}, is here repositioned as a universal model for diffusional earthquakes (Fig. 3b, c). The internal process driving the diffusion is Brownian motion, namely migration proceeding at a constant speed but in randomly varying directions. %Thus, asserting that the spatial average of the rupture is diffusive is phenomenologically equivalent to positing that individual rupture fronts migrate randomly. 
The Brownian rupture model thus describes the rupture as a superposition of rupture elements migrating via Brownian motion, which represents a diffusional rupture cascade propagating at a constant speed while erratically altering its course (Fig. 3b), contrasting with the standard cascading rupture model macroscopically fixing its speed and course~\cite{ellsworth1995seismic}. 
In the Brownian model, the rupture zone is represented by the coverage area $S_{\rm true}$ of trajectories from multiple Brownian walkers, yielding the seismic moment $M_0=\mu S_{\rm true} \delta_*$ with a characteristic average slip $\delta_*$. 
Because spatial overlaps between trajectories, leading to occasional asperity re-ruptures due to stress re-build up~\cite{ide2019two}, make the seismicity area a nonlinear function of trajectories, the superposition principle holds only in the asymptotics. The true active area $S_{\rm true}$ is therefore smaller than the area $S$ fitted by the diffusive envelope $S=Dt$ ($S_{\rm true}<S$). 
The first order solution of $S_{\rm true}$ is obtained by simplifying this path-overlapping effect, omitting the recounts of overlapped trajectories. Under this approximation, the seismicity area of a Brownian rupture, and thus the moment release approximately proportional to it, corresponds to the number of distinct sites visited by random walkers on a lattice (Fig. 3c). Its analytical expression $g$ is given by~\cite{weiss1983random,larralde1992territory}:
\begin{equation}
M_0 = \mu \delta_* g(S) \simeq \mu \delta_* S\min (S/S_*,1).
\end{equation}
Here, $\delta_*$ and $S_*$ represent fitting parameters of the scaling function (see Methods). 
This functional form explains the observed mean-slip upper bound as an underlying uniform constant slip $\delta_*$ of the Brownian rupture, reproducing the simulated scaling of the original proposal~\cite{ide2008brownian}. The Brownian rupture model successfully fits the transient phase leading to the constant-slip slope spanned by VLFEs, supported by the description that a VLFE is an ensemble of LFEs~\cite{shelly2007non} which follow $M\propto t^3$ as in ordinary earthquakes~\cite{supino2020self}. Accordingly, this theoretical model unifies diffusional earthquakes tracing similar $M_0$-$S$ trajectories, excepting LFEs and initial swarm activities potentially representing the elements of diffusional activities (Fig. 3c).

The present findings offer a perspective on the universal 3 MPa stress drop for ordinary earthquakes~\cite{allmann2009global}: ruptures arrest before accumulating centimeter-order slip unless this stress-release level is achieved. The characteristic slip $\delta_* \sim 1$ cm for diffusional earthquakes is smaller than the critical slip-weakening distance expected for large events, suggesting that premature rupture termination is their character. In Brownian rupture models, the rupture unit coarse-grained into the lattice represents the typical distance to change the migration direction, effectively captured by $\sqrt{S_*} \sim 3$ km. 
Because the drivers of diffusional earthquakes likely vary, such as fluid intrusion~\cite{shapiro2009scaling,mcgarr2014maximum} and triggered aseismic slip~\cite{vidale2006survey,bletery2017characteristics,saez2023post}, their common scaling relation implies a universal environment underlying their occurrence. 

Systematic deviations from this scaling, which otherwise prohibits large cascade constituents with slips exceeding tens of centimeters, might signal transitions toward large, hazardous earthquakes~\cite{yoshida2023upward,okuwaki2024multiplex}. However, we note that the similarity hypothesis for diffusional earthquakes, based on the superposition of small constituents, remains debated, particularly regarding SSEs~\cite{michel2019similar,tan2020connecting}. The Brownian scaling might also be debated as an apparent coincidence. 
If fitted with standard circular cracks, diffusional earthquakes would imply unusually low stress drops ($\sim$30 kPa), though our cumulative slip pattern supports the constant-slip pulse. Elucidating why diffusional earthquakes behave as Brownian ruptures is still an issue, which is in parallel to understanding why ordinary earthquakes fit circular cracks.

\section*{Materials and Methods}
\subsection*{Moment release evaluation of cataloged events}

We used hypocenter data for the Yamagata-Fukushima-border and Sendai-Okura earthquake swarms relocated in previous studies~\cite{yoshida2018hypocenter,yoshida2018sendai}. The former includes events occurring within 800 days after the Tohoku-Oki earthquake (March 11, 2011) within the spatial range of 139.885${}^\circ$--140.065${}^\circ$E and 37.63${}^\circ$--37.85${}^\circ$N. The latter spans from the Tohoku-Oki earthquake to December 31, 2015, within the spatial range of 140.67${}^\circ$--140.75${}^\circ$E and 38.29${}^\circ$--38.35${}^\circ$N. The coordinate origin was set at the initiation point of the swarm activity, which we evaluated as the average hypocenter of the first five events. 

We identified the JMA magnitude with the moment magnitude. Applying an empirical conversion formula between the JMA and moment magnitudes~\cite{edwards2009comparative} substantially modifies the frequency distribution, primarily for events smaller than magnitude 1 (Fig. S2a). However, this results in a difference of roughly a factor of two in the total moment release of the swarm (Fig. S2b), which does not affect the discussion in the main text. Regarding the 5\% outliers removed from the analysis, their moment release occasionally lowers the cumulative magnitude by nearly 1 at certain times during the initial 1-day period, when the number of events is relatively small. Thereafter, however, their contribution becomes negligible compared to the total moment release (Fig. S2c).

\subsection*{Geometrical outlier removal}

Because the rupture areas of individual events are negligibly small compared to the entire seismicity area, evaluating the active area of the earthquake swarm, which is the primary focus of our scaling analysis, reduces to calculating the effective area occupied by the hypocenter point cloud. The area evaluation requires extracting the spatial orientation of the main geometric structure (fault plane) while mitigating overestimations of the envelope caused by a small number of scattered events, such as poorly located isolated hypocenters or background seismicity. Therefore, we excluded outliers corresponding to 5\% of the total events prior to the area calculation.
In this study, we utilized the Isolation Forest because it allows the outlier exclusion ratio to be set as an explicit hyperparameter, referred to as the contamination ratio.

We verified the validity of this exclusion rate a posteriori by examining its parameter dependence and susceptibility to algorithm alteration (Fig. S3). When varying the contamination ratio in the Isolation Forest, the 3D volume and 2D surface area derived from the convex hull of the valid point cloud, as well as the convex and concave hull areas in the 2D coordinate system projected onto the $X_1$-$X_2$ plane, exhibit L-shaped trends against the exclusion rate (Figs. S3a1 and S3a2). The adopted exclusion rate of 5\% is located at the corner of these L-shaped curves. Furthermore, we evaluated the inner products ($|X_i \cdot Y_i|$) between the principal axis vectors of the fault strike ($Y_1$), dip ($Y_2$), and normal ($Y_3$), extracted via principal component analysis (PCA) and their respective reference axes ($X_i$, for 5\% outlier removal) adopted in the main text. Across a broad parameter range, these inner products are stable with an error of less than 1\%, indicating that the spatial orientation of the extracted fault plane is largely independent of the outlier removal process (Fig. S3a3). When outliers were removed using the HDBSCAN algorithm described later, we obtained results similar to those achieved by removing the 5\% Isolation Forest outliers across a wide parameter range (Fig. S3b).

\subsection*{Concave-envelope determination via maximum negentropy}

Because the active area of an earthquake swarm contains fractal-like voids, a simple convex hull overestimates the effective area. To mitigate this overestimation, we adopted a concave hull. Its concavity hyperparameter (a fraction of the difference between the longest and shortest edge lengths in the Delaunay Triangulation, here denoted by convexity\_parameter) was determined by maximizing the distance from a normal distribution (negentropy) for the logarithmic edge-length distribution of the polygons forming the envelope (Fig. S4). This measure represents a structural deviation from Gaussian disorder realized under the central limit theorem, useful for capturing the clustered nature of seismicity discussed in the main text.

The micro-scale edge-length distribution formed by the Delaunay triangulation of the original point cloud, representing inter-event distances, follows a log-normal distribution (Fig. S4a, Original Delaunay Edges). Conversely, the log-edge-length distribution of the macroscopic convex hull enclosing the entire system (convexity\_parameter = 1.0) is represented by delta functions due to the limited number of edges, yet it similarly possesses characteristic distances (Fig. S4a, Convex). When varying the convexity\_parameter, the higher-order cumulants of the envelope's log-edge-length distribution identify signals intermediate between these two limits (Fig. S4b). To stably estimate the cumulants and prevent the overvaluation of specific edges, we computed the ensemble expected values and standard errors using a subsampling method without replacement (extracting 80\% of all edges without duplication for each parameter, with 500 trials).

The negentropy was evaluated via the Edgeworth expansion based on the standardized cumulants $\kappa_i$, which is an asymptotic expansion with respect to sample size. Because the fifth-order cumulant exhibits an error of nearly 100\% around the negentropy peak, the Edgeworth expansion is truncated at the standard, second-order expansion: $J=\kappa_3^2/12 + \kappa_4^2/48 + 7\kappa_3^4/48 -\kappa_3^2\kappa_4/8 + \dots$~\cite{comon1994independent}. The second-order negentropy exhibits a nearly unimodal distribution (Fig. S4c), and we adopted convexity\_parameter = 0.15 at this global maximum in the main text. (For the Sendai-Okura swarm, where a similar analysis was performed, the number of detected events is in the thousands, making the cumulant estimation less stable than that for the Yamagata-Fukushima border. Therefore, in the hyperparameter selection, convexity\_parameter values resulting in fewer than 50 edges were excluded from the evaluation interval. When the number of edges exceeded this threshold, the distribution profile for the Sendai-Okura swarm was also unimodal.) Unlike the Gaussian log-edge-length distributions of simple inter-event distances and the convex hull, the envelope shape obtained through negentropy maximization extracts non-Gaussian cluster boundaries that capture the differentiation of the source region.

\subsection*{Density-based hierarchical clustering via entropy optimization}
To isolate the spatiotemporal evolution of individual rupture segments within the earthquake swarm, we performed clustering based on spatial coordinates. As previously described, the micro-scale inter-event distances in this system follow a log-normal distribution. Consequently, when employing methods based on a single distance threshold (e.g., DBSCAN~\cite{ester1996density}), the long tail of the probability distribution causes the coverage area to increase gradually with the threshold (Fig. S4a), making it difficult to stably detect the intermediate structure identified earlier. To simultaneously classify the majority of spatially cohesive small-to-medium-sized clusters and the large clusters probabilistically linked through the log-normal tail, we adopted the HDBSCAN~\cite{mcinnes2017hdbscan} algorithm, which accommodates hierarchical density structures. 

The minimum cluster size ($N_{\min}$), a hyperparameter in HDBSCAN, was determined based on the variation of the Shannon entropy (partition entropy) of the membership probabilities. When varying $N_{\min}$, there exists a plateau region where the number of extracted clusters and their total area stabilize; within this range, the partition entropy hits a flat local minimum (Fig. S4b). 
Because the entropy trivially reduces to zero if all events merge into a single overarching cluster, this flat zone is the local minimum, representing a structurally meaningful optimal state. 
This local-minimum-entropy state avoids both overfitting to specific clusters and excessive merging, corresponding to the classification that most meaningfully compresses the information.

The sum of the convex hull areas of these individual sub-clusters is consistent with the concave envelope area identified independently via negentropy maximization (Fig. 1c). This demonstrates that the complementary approaches of macroscopic structured envelope extraction (maximum negentropy) and microscopic optimal event classification (local-minimum entropy) yield a consistent effective area.

\subsection*{Compilation of reported diffusional earthquakes}

To compare the earthquake swarms analyzed in this study with various diffusional earthquakes (Fig. 2), we utilized event catalogs from previous studies~\cite{michel2019similar,bletery2017characteristics,danre2022prevalence,supino2020self,ide2018seismic}.

Duration estimates were directly adopted from the reported values in the catalogs where available. For the LFE catalog~\cite{supino2020self}, which lists corner frequencies $f_{\rm c}$ rather than durations $t$, we converted $f_{\rm c}$ to $t$ as $t = 1 / (\pi f_{\rm c})$ based on the spectral-domain envelope of a boxcar source time function~\cite{ide2007scaling}.

The plotted seismicity areas represent the spatial extents of the events, evaluated in accordance with the rupture geometries assumed in their original source models; we use them as proxies for the seismicity area. For natural earthquake swarms and injection-induced seismicity~\cite{danre2022prevalence}, we directly adopted the areas estimated via standard circular crack models.
The areas of SSFs~\cite{bletery2017characteristics} were evaluated as rectangular ruptures ($S = LW$) using the length $L$ (representing the distance along the best projection axis) and the width $W$ (the 95\% confidence interval orthogonal to the best projection axis). 
For SSEs~\cite{michel2019similar}, we assumed elliptical ruptures to evaluate the seismicity area ($S = \pi R_1 R_2$) based on the cataloged length (the distance between the northern and southern intersections of the rupture's outline with the mean strike line) and width (twice the mean distance between the rupture's outline and the mean strike line). Because the catalog's maximum and minimum bounds for the SSE moment and duration differ negligibly on a logarithmic scale, we adopted their average values. We remark that while the rupture area exceeding the slip threshold of these SSEs scales with the 3/5 to 2/3 power of the seismic moment~\cite{michel2019similar}, their elliptical envelope, which aligns more closely with the definition of seismicity area used here, consequently enhances the linearity between area and moment, yielding a power-law exponent of $\sim$ 3/4.
For LFEs~\cite{supino2020self} and VLFEs~\cite{ide2018seismic}, whose catalogs primarily provide the temporal extents of the events, we evaluated circular seismicity areas ($S = \pi R^2$) using empirical conversions of the duration $t$. The effective radius $R$ was estimated as $R = v_{\rm r} t$ with a rupture speed of $v_{\rm r} = 0.7$ km/sec~\cite{thomas2016constraints} for LFEs, and as $R = \sqrt{Dt}$ with a diffusion coefficient of $D = 10^4$ m$^2$/sec~\cite{ide2018seismic} for VLFEs.

The scaling asymptotics for ordinary and slow earthquakes used as reference lines are based on a previous study~\cite{ide2007scaling}. For the lower moment-rate band for fluid-related seismicity, we referred to the $M_0 \propto t$ scaling from a previous study~\cite{danre2024parallel} and set its intercept to serve as a support line for the Yamagata-Fukushima border swarm. Calculations for the 3 MPa stress drop and 1 cm slip distance reference lines assumed a circular crack and a rigidity of 40 GPa.

\subsection*{Cumulative slip computations based on the circular crack model and resolution analysis}

To evaluate the spatial distribution of the cumulative slip distance, we modeled each earthquake event as a circular crack with a constant stress drop of 3 MPa. The stress drops estimated from the corner frequencies of this swarm are consistent with this assumption~\cite{yoshida2017temporal,yoshida2018hypocenter}. We projected the hypocenter distribution onto a fault plane defined by the strike ($X_1$: N30$^{\circ}$E) and up-dip ($X_2$: S60$^{\circ}$E, 15$^{\circ}$ up-dip) axes, where a circular slip distribution centered at the hypocenter determines the slip pattern of each event. In the circular crack model, the radius $R$ is calculated from $M_0=(16/7) \Delta \tau R^3$, and the slip $\delta$ inside the crack at a coordinate ${\bf x}$ originating from the hypocenter is given by $\delta = [3M_0/(2\mu S)] \sqrt{1 - |{\bf x}|^2/R^2}$. 
Although this simple kinematic superposition neglects complex stress interactions among neighboring events, we adopted it as a first-order approximation to evaluate the macroscopic cumulative slip. As in the main text, we assumed a rigidity of 40 GPa.

For the computation, the fault plane is discretized into a grid. The events of the Yamagata-Fukushima border swarm analyzed in the main text form a point cloud of approximately 30,000 events distributed over a 15--20 km $\times$ 3--5 km region (Fig. 3a), with an average inter-event interval of roughly 50 m. Given this spatial density, we adopted a grid size of 100 m in the main text, which corresponds to approximately twice the average interval.

We simplified the slip computation for each event based on the ratio of the crack diameter to the grid size. If the crack diameter is smaller than the grid size, we applied a point-source approximation: the event's entire seismic moment was assigned to the single grid cell containing its hypocenter, increasing the average slip on that cell by the seismic moment divided by the product of the rigidity and the grid area. On the other hand, if the crack diameter exceeds the grid size, we employed a point-collocation approximation: the slip on multiple grid cells was evaluated based on the distance from the crack center to each grid center.

To confirm the robustness of the spatial patterns against the approximations described above, we compared the results by increasing the grid size by factors of two (200 m) and four (400 m) (Fig. S6). The overall morphology formed by the overlapping slip flakes is insensitive to the resolution. Although a larger grid size enhances the spatial smoothing effect, the average slip distance within the grid cells where slip occurs (i.e., cells with nonzero slip) does not significantly depend on the resolution (2.0 cm, 1.5 cm, and 1.0 cm for grid sizes of 100 m, 200 m, and 400 m, respectively). 
The logarithmic slip distributions show that the broad peak around the centimeter scale ($\log_{10}(\text{slip [cm]}) \approx 0$) smooths out upon coarse-graining with larger grid sizes (Fig. S6c).

Note that the average slip distance $\bar{D}$ evaluated here represents the mean slip distance over the rupture area in this coarse-grained model. This quantity is distinct from the mean slip over the enveloped seismicity area ($M_0/S$) discussed in the main text, resulting from the difference between $S$ and $S_{\rm true}$ mentioned in the main text. However, because the seismicity densely fills the active region during the later stages of the swarm, these two values may be considered practically comparable. Indeed, the estimated $\bar{D}$ on the order of centimeters is consistent with the asymptotic value of $M_0/S$ ($\sim$1 cm) derived from the macroscopic scaling. Furthermore, the effective area estimated from the total seismic moment of the swarm ($M_{\text{w}} \approx 5.1$) and the average slip $\bar{D}$ of 1–2 cm ranges from 70 to 140 km$^2$, consistent with the enveloped area evaluated in the main text. Because the swarm exhibits fractal-like clustering with voids, the apparent geometric area increases with the coarse-graining scale. The 100 m grid size is close to the lower limit defined by the average inter-event interval ($\sim$50 m), corresponding to the extreme case of a concave hull (convexity\_parameter $\to 0$). The effective rupture area evaluated at this finely resolved scale thus serves as a lower bound of the enveloped seismicity area.

\subsection*{Simulation of Brownian rupture and conversion to scaling relations}

Implementations of Brownian rupture can be broadly classified into two types: continuum models based on stochastic partial differential equations for the rupture radius~\cite{ide2008brownian} and discrete models based on cellular automata~\cite{ide2019two}. In this study, we followed the latter approach formulated on a two-dimensional square lattice. Specifically, in the initial state ($t=0$), $N_{\rm walkers}$ random walkers are placed at the origin. At each time step, each walker independently transitions to one of the adjacent sites (up, down, left, or right) with equal probability (1/4). As mentioned in the main text, re-ruptures of sites are not accounted for in this study. We do not introduce a complex graph structure analogous to a microscopic asperity network. Investigating the dependence on these structural factors is left for future studies; here, we limit our scope to characterizing the behavior within the simple model employed, leaving aside further complexities not incorporated in previous studies.

The simulation examples are shown in Fig. S7. To show the expected scaling behavior, ensemble averaging was performed for each $N_{\rm walkers}$ using an adaptive number of ensembles ($N_{\rm ens} \approx 600 / \sqrt{N_{\rm walkers}}$). As shown in Fig. S7, the time evolution of the cumulative number of visited sites, representing the discrete true rupture area $A_{\rm true}$, is characterized by two distinct scaling regimes. In the early stage of time evolution, the ruptured area grows almost ballistically as walkers fill the rupture zone densely. In the long-time regime, the walkers become sparse, leading to a diffusive growth that asymptotically approaches $A_{\rm true} \simeq Dt$. To systematically determine the crossover time $t_*$, we first estimate the macroscopic diffusion coefficient $D$ by fitting the long-time regime ($300 \leq t \leq 10,000$) with a linear regression in the log-log space. Based on the continuum limit where a dense circular expansion ($A_{\rm true} = \pi t^2$) intersects with the diffusional growth ($A_{\rm true} = Dt$), the theoretical crossover is initially guessed as $t \approx D/\pi$. Using this geometric estimation as an upper bound, we perform a second log-log linear fit for the early ballistic regime ($1 \leq t \leq D/\pi$). Finally, the actual crossover time $t_*$ is determined as the intersection of these two empirically fitted scaling laws. The resultant $t_*$ meets this first-order evaluation, yielding $D/t_* \approx 2$ in the simulated range of $N_{\rm walkers}$. This prefactor of roughly 2 (rather than $\pi$) is a natural geometric consequence of the lattice random walk, where the lattice determines the equidistance space~\cite{weiss1983random}; for the squared lattice, the grid, the L1 norm equidistance space results in a square-shaped rupture front (i.e., $A_{\rm true} = 2 t^2$) in the ballistic regime. We emphasize that the scaling behavior itself is here unaffected by this lattice configuration, preserving its physical validity in the continuum limit.

%It is evident that the crossover time $t_*$ is an increasing function of $N_{\rm walkers}$ because the system continuously exhibits ballistic growth in the limit of $N_{\rm walkers} \to \infty$. However, as long as a finite $N_{\rm walkers}$ is considered, the asymptotic discrete diffusion coefficient $D$ increases only logarithmically with $N_{\rm walkers}$, making its dependence fairly weak. 

To compare the behavior of the obtained discrete model with the macroscopic scaling of actual diffusive earthquakes, we translated the simulation results for $N_{\rm walkers} = 10^5$ into physical quantities in continuous space, as shown in Fig. 3c. The mapping procedure is structured through the following steps. First, in the simulation, the number of ruptured sites represents the discrete true rupture area $A_{\rm true}$. It follows a scaling function with respect to the time step $t$, transitioning from an early ballistic expansion ($A_{\rm true} \propto t^2$) to asymptotic diffusion ($A_{\rm true} \propto t$), expressed as $A_{\rm true} \simeq Dt \min(t/t_*, 1)$. Second, by defining the discrete area as its asymptotic scaling $A = Dt$, the explicit time dependence is eliminated, yielding $A_{\rm true} \simeq A \min(A/A_*, 1)$, separated by the discrete crossover area $A_* = Dt_*$. Third, we map this discrete geometric relation onto the continuous physical space using an area conversion factor $\alpha$, obtaining the macroscopic seismicity area $S = \alpha A$ and the physical true rupture area $S_{\rm true} = \alpha A_{\rm true}$. By assigning a uniform constant slip $\delta_*$ to the ruptured sites, the seismic moment is evaluated as $M_0 = \mu \delta_* S_{\rm true}$, assuming a rigidity of $\mu = 40$ GPa. Consequently, the model yields the scaling function $M_0 \simeq m_* S \min(S/S_*, 1)$, where $m_*:=\mu \bar{\delta}$, and $S_* = \alpha A_*$ is the crossover area. Comparing this formulation with the scaling relation for actual diffusional earthquakes demonstrates that the upper bound of the slip averaged over the envelope $\bar{\delta}$ is identical to the characteristic slip distance $\delta_*$ of the rupture area in the Brownian rupture model. The conversion factor $\alpha$ was explicitly determined by equating the crossover area $S_*$ to the intersection of the VLFE band and the constant slip line for $\delta_* = 1$ cm, which we approximated to be $10^7$ m$^2$.

Crucially, the asymptotic shape of the Brownian scaling solution is entirely governed by $D$ and $S_*$. The total number of random walkers $N_{\rm walkers}$ merely adjusts the duration of the ballistic regime ($S \ll S_*$). This $N_{\rm walkers}$ dependence relates to the question of how many LFEs superimpose to act as a VLFE, although a detailed quantitative evaluation of this aspect is beyond our current scope. We simply used $N_{\rm walkers} = 10^5$ given the weak size dependence of $t_* \propto \log N_{\rm walkers}$ (Fig. S7b), as a sufficient size to reasonably cover the observed ballistic-scaling range; this dependence reflects the fact that the ballistic regime lasts forever for an infinite number of walkers. Since the ballistic range produced by $N_{\rm walkers} = 10^5$ covers the observed scale of VLFEs, this population size is within the size range to fit the model to the observations. Because the discrete initial condition treats the first event as nominally instantaneous, its physical rise time is unresolved at the origin time. To purposely mitigate this issue, we extrapolated the data back to the single-asperity limit by extending the slope of the early ballistic trajectory ($t \le 2$) on the log-log scale. This extrapolation, indicated by the dotted line in Fig. 3c, aligns the early-stage scaling with the trajectory expected for larger $N_{\rm walkers}$. As emphasized in the main text, since Fig. 3c is a spatial plot relating seismic moment and ruptured area, the intrinsic time constant or time scale is irrelevant to the shape of the plot.

% Research Articles and Reviews split the text into sections using headings
% Use a short (up 6 words) descriptive phrase, not generic 'Results' or 'Conclusions'
% Most other formats do not have headings, see the journal instructions to authors for details
%\section*{An example heading}
%\subsection*{An example heading}
%Research Articles and Reviews use headings to split the main text into sections; most other formats do not have headings.

% If your text is very short you might need to uncomment the following line to avoid
% layout problems with the figures and tables.
\newpage

%%%%%%%%%%%%%%%% MAIN TEXT FIGURES %%%%%%%%%%%%%%%

%\begin{figure} % Do NOT use \begin{figure*}
%	\centering
%	\includegraphics[width=0.6\textwidth]{example_figure} % for an image file named example_figure.*
	% Pick an appropriate width - in print, figures are usually one or two columns wide, which can
	% be approximated by 0.3\textwidth or 0.6\textwidth respectively. Use appropriate label sizes.
	% Captions go below figures

\begin{figure*}
\centering
 \includegraphics[width=150mm]{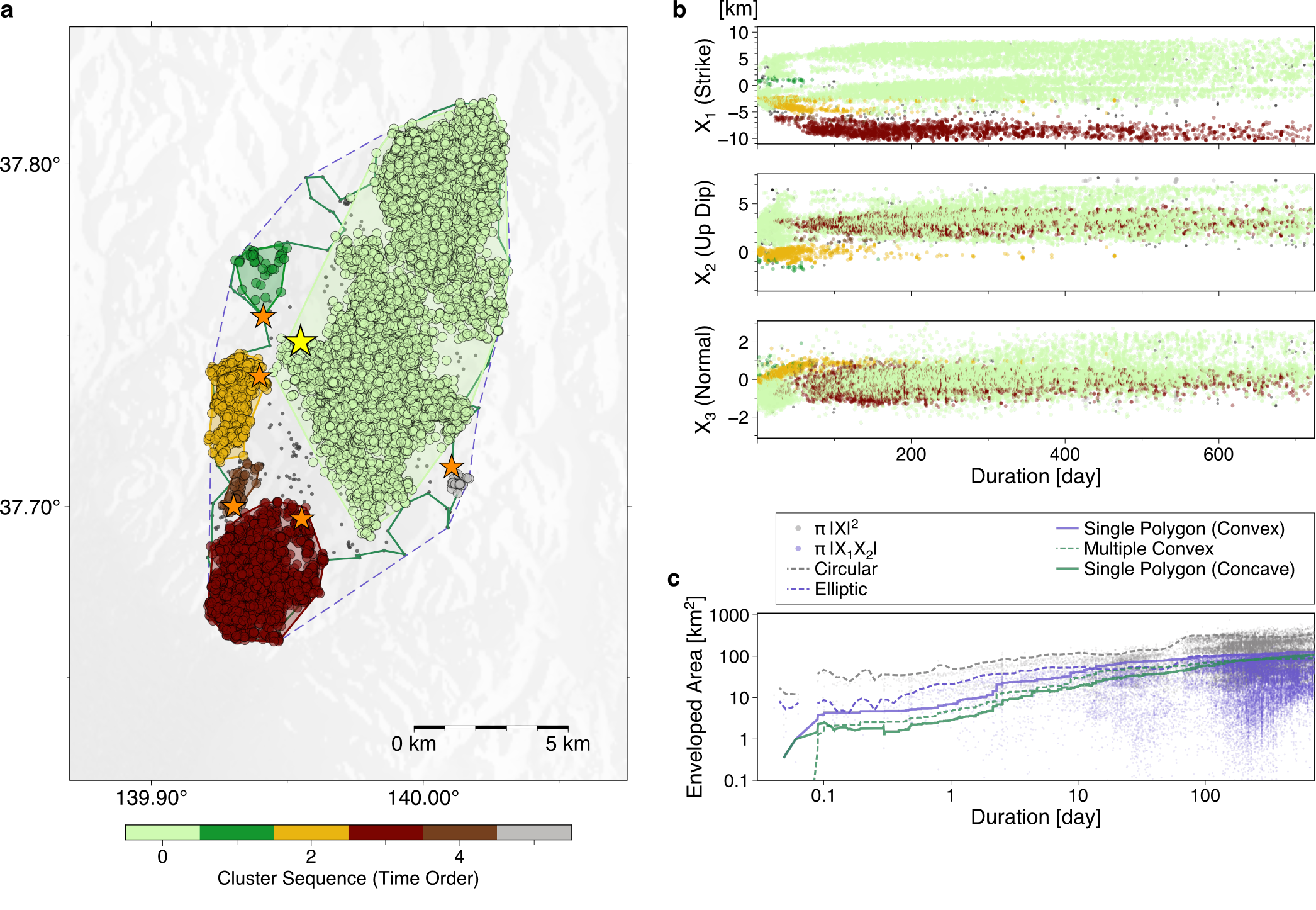}
 \caption{
Spatiotemporal evolution of the Yamagata-Fukushima-border earthquake swarm. (a) The relocated hypocenters. Star symbols indicate the initiation points of the entire activity (yellow) and of each cluster (orange). Each cluster is colored according to its onset time, and unclustered points are shown in black. Removed 5\% outliers are unplotted. Lines represent the envelopes of the convex hull for the entire points (dashed black), the concave hull for the entire points (solid green), and the convex hull of each cluster (solid lines, colored correspondingly). 
(b) Spatiotemporal evolution along the first ($X_1$), second ($X_2$), and third ($X_3$) principal components, which correspond to the strike, dip, and normal, respectively. The origin is set at the activity initiation point. The color scheme follows Fig. 1a. 
(c) Area evaluation based on the concave hull in comparison with other geometric metrics. The background scatter plots show the evolution of the isotropic ($\pi |X|^2$, gray) and anisotropic distances ($\pi |X_1 X_2|$, purple) from the initiation point. Their corresponding $2\sigma$ envelopes (circular and elliptical) are plotted, along with the areas of the concave (solid purple) and convex (solid green) hulls for filtered events. The area sum of the convex hulls for individual clusters (dotted green) is also indicated.
}
 \label{fig:1}
\end{figure*}

\begin{figure*}
\centering
 \includegraphics[width=150mm]{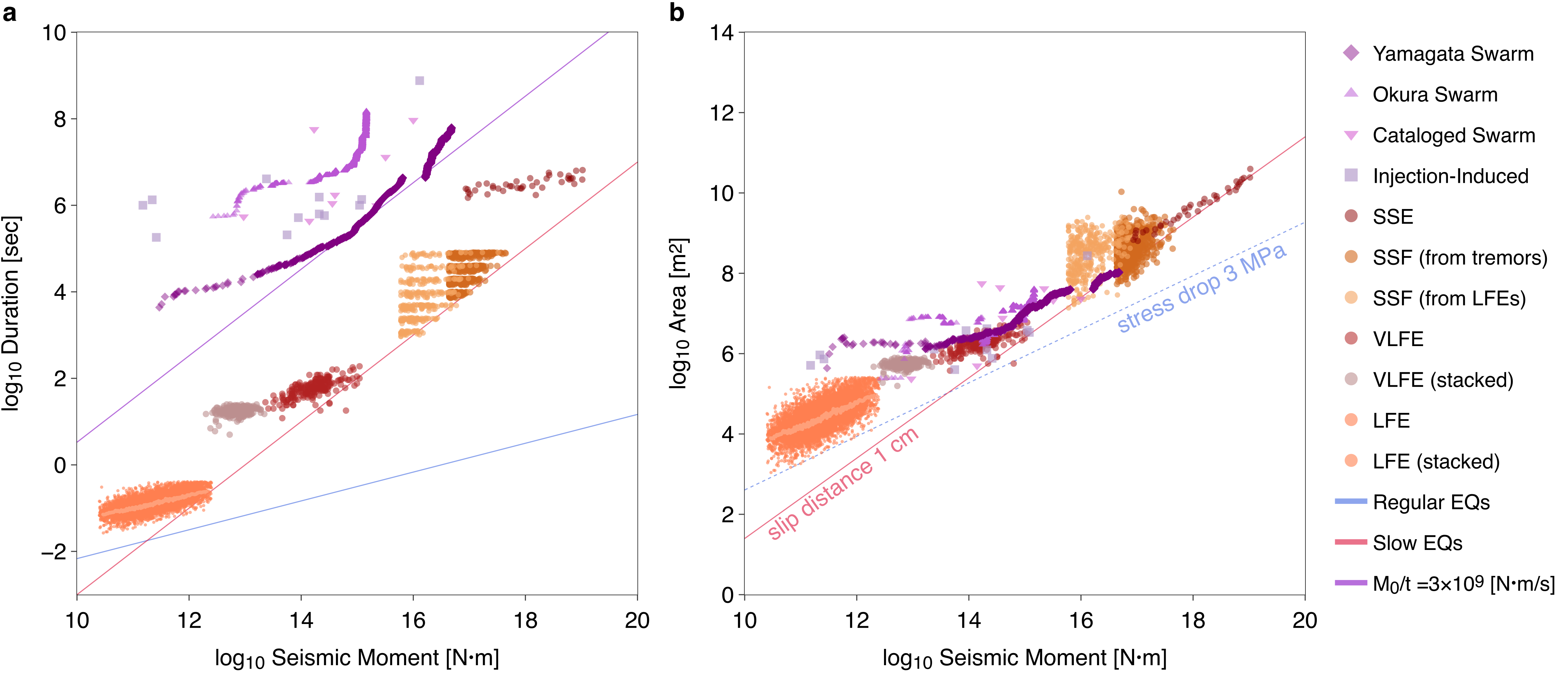}
 \caption{
 Seismic moment of diffusional earthquakes versus duration (a) and seismicity area (b). The analyzed earthquake swarms (Yamagata and Okura) are plotted alongside reported natural swarms, injection-induced seismicity, and various slow earthquake phenomena, which include low-frequency earthquakes (LFEs), very low-frequency earthquakes (VLFEs), slow slip fronts (SSFs) derived from tremors and LFEs, and slow slip events (SSEs). 
 Reference lines in Fig. 2a denote the asymptotic scalings for regular earthquakes ($M_0\propto t^3$, eq. 3, blue) and slow earthquakes ($M_0\propto t$, eq. 4, red) and the lower moment-rate band for fluid-related seismicity (purple). Reference lines in Fig. 2b denote asymptotics assuming the constant stress drop of 3 MPa (blue) and the constant slip of 1 cm (red). 
}
 \label{fig:2}
\end{figure*}

\begin{figure*}
\centering
 \includegraphics[width=150mm]{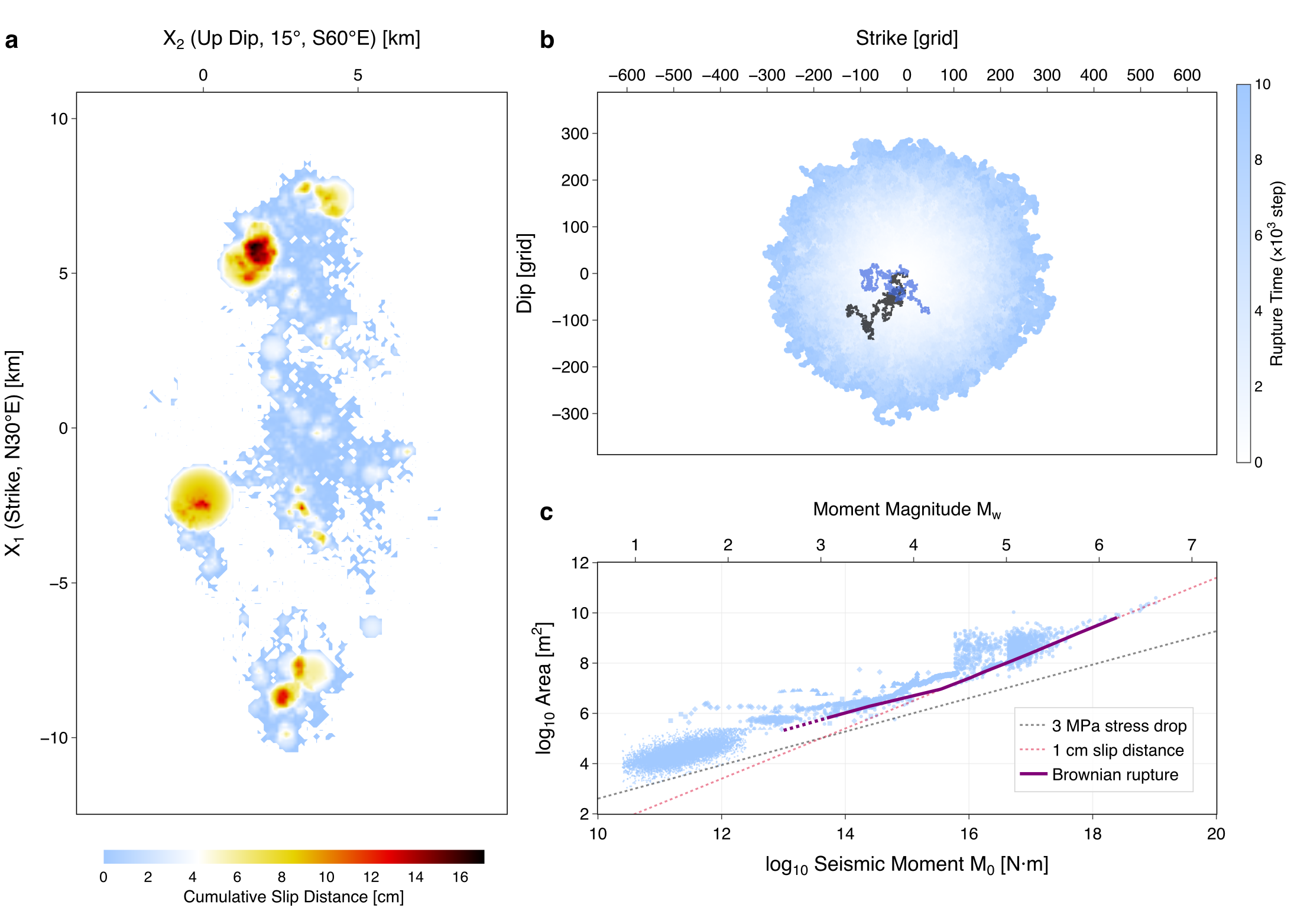}
 \caption{
 Semiphysical modeling of the Yamagata-Fukushima-border earthquake swarm and diffusional earthquakes.
 (a) Collective crack model of the Yamagata-Fukushima-border earthquake swarm. Each event is modeled as a circular crack with a constant stress drop of 3 MPa, projected onto an effective fault plane spanned by $X_1$ (strike) and $X_2$ (up-dip) axes. The colors indicate the resulting cumulative slip distance.
 (b) Simulated Brownian rupture model. 
 The time evolution of seismicity is computed using $10^5$ random walkers, each of which represents a stochastic rupture cascade. The color gradient indicates the first visit time of the random walkers, corresponding to the time of rupture. The lines trace two representative random walks.
 (c) Moment-area scaling of the Brownian rupture model. The scaling curve converted from the simulation (purple line) is superimposed on the compiled observational data of various slow earthquakes and fluid-related seismicity. The dotted segment extrapolates the rise time of the nominally instantaneous initial rupture at the origin grid, configured to match the trajectory for a larger number of walkers.
 Theoretical scaling relations for a constant 3 MPa stress drop (gray dotted line) and a constant 1 cm slip distance (red dotted line) are also plotted.
}
 \label{fig:3}
\end{figure*}

%%%%%%%%%%%%%%%% MAIN TEXT TABLES %%%%%%%%%%%%%%%
%
%\begin{table} % Do NOT use \begin{table*}
%	\centering
%	% Captions go above tables
%	\caption{\textbf{All captions must start with a short bold sentence, acting as a title.}
%		Then explain what is being listed in the table, the meaning of each column etc.
%		Captions are placed above tables.}
%	\label{tab:example} % give each table a logical label name
%	
%	\begin{tabular}{lccc} % four columns, alignment for each
%		\\
%		\hline
%		Sample & $A$ & $B$ & $C$\\
%		 & (unit) & (unit) & (unit)\\
%		\hline
%		First & 1 & 2 & 3\\
%		Second & 4 & 6 & 8\\
%		Third & 5 & 7 & 9\\
%		\hline
%	\end{tabular}
%\end{table}
%
%
%
%%%%%%%%%%%%%%%% REFERENCES %%%%%%%%%%%%%%%

\clearpage % Clear all remaining figures and tables then start a new page

% The list of references goes after the main text and before the acknowledgements
% When preparing an initial submission, we recommend you use BibTeX, like this:
%
\bibliography{diffusional_earthquakes}
\bibliographystyle{sciencemag}

% After the paper has completed peer review and been revised ready for acceptance,
% you should comment out the lines above and copy-paste the contents of your .bbl
% file here instead. This will help ensure that our conversion software works correctly.
% Remember to re-run BibTeX first - check the timestamp!
%
% Example of the first three entries copy-pasted from science_template.bbl:
%
%\begin{thebibliography}{1}
%
%\bibitem{example}
%A.~N. {Author}, An example reference. \emph{Journal of Improbable Research}
%  \textbf{1}, 67 (2020).
%
%\bibitem{example2}
%F.~M. {Surname}, S.~{Author}, A second example. \emph{Interesting Research
%  Letters} \textbf{32}, 897 (2019).
%
%\bibitem{example_preprint}
%P.~{One}, P.~{Two}, P.~{Three}, {An unpublished preprint}. \emph{preprint}
%  (2021), arXiv:2101.12345.
%
%\end{thebibliography}

%%%%%%%%%%%%%%%% ACKNOWLEDGEMENTS %%%%%%%%%%%%%%%

\section*{Acknowledgments}
We thank Yoshihiro Kaneko and Tomoaki Nishikawa for their valuable comments.

\paragraph*{Funding:}
This study was supported by JSPS KAKENHI Grants (23K19082 and 21H05206).

\paragraph*{Author contributions:}
D.S.: Conceptualization, Formal analysis, Funding acquisition, Methodology, Visualization, Writing - original draft.
K.Y.: Data curation, Resources, Validation, Writing - review \& editing.

\paragraph*{Competing interests:}
There are no competing interests to declare.

\paragraph*{Data and materials availability:}
The earthquake catalog analyzed in this study is available from the Japan Meteorological Agency (\url{https://www.data.jma.go.jp/suishin/catalogue/quake.html}). 
Compiled scaling data for other seismic events were sourced from the literature~\cite{supino2020self,ide2018seismic,bletery2017characteristics,michel2019similar,danre2022prevalence}.
%Compiled data are available from Supino et al. (2020)~\cite{supino2020self} for tremors, Ide and Maury (2018)~\cite{ide2018seismic} for VLFEs, Bletery et al. (2017)~\cite{bletery2017characteristics} for SSFs, Michel et al. (2019)~\cite{michel2019similar} for SSEs, and Danre et al. (2022)~\cite{danre2022prevalence} for natural swarms and injection-induced seismicity.
All statistical and geometrical computations, including hull constructions~\cite{geos2025geos}, point-cloud clustering~\cite{liu2008isolation,ester1996density,mcinnes2017hdbscan}, and entropy-based optimizations~\cite{comon1994independent}, were implemented using standard open-source Python libraries.%, primarily scikit-learn~\cite{pedregosa2011scikit}.
%Concave hull computations were implemented using the Geometry Engine Open Source (GEOS) library~\cite{geos2025geos}. %and its Python wrapper, Shapely~\cite{gillies2007shapely}. For outlier removal and clustering, we utilized the Isolation Forest~\cite{liu2008isolation} and HDBSCAN~\cite{mcinnes2017hdbscan} algorithms, respectively, via the open-source Python library scikit-learn~\cite{pedregosa2011scikit}; DBSCAN~\cite{ester1996density} was also employed to verify the clustering results. The entropy calculations for hyperparameter optimization utilized the built-in functions of scikit-learn, while the negentropy calculations were based on the Edgeworth expansion~\cite{comon1994independent}.

%%%%%%%%%%%%%%%% SUPPLEMENT LIST %%%%%%%%%%%%%%%

% List the contents of your Supplementary Materials, including the numbers of any
% supplementary figures, tables, external data files etc. and any references that are
% cited only in the supplement. In this example, refs. 7-8 are cited only in the supplement.
% Fill out your numbers accordingly and delete any lines that aren't applicable.
\subsection*{Supplementary materials}
%Materials and Methods\\
%Supplementary Text\\
Figs. S1 to S7\\%Figs. S1 and S2\\
%Tables S1 to S3%\\
%References \textit{(7-\arabic{enumiv})}\\ % automatically fills out the last reference number
% (filling out the other numbers automatically is possible but fiddly and liable to break)
%Movie S1\\
%Data S1

%%%%%%%%%%%%%%%% END OF MAIN TEXT %%%%%%%%%%%%%%%

\newpage

%%%%%%%%%%%%%%%% START OF SUPPLEMENT %%%%%%%%%%%%%%%

% Figures, tables, equations and pages in the supplement are numbered S1, S2 etc.
\renewcommand{\thefigure}{S\arabic{figure}}
\renewcommand{\thetable}{S\arabic{table}}
\renewcommand{\theequation}{S\arabic{equation}}
\renewcommand{\thepage}{S\arabic{page}}
\setcounter{figure}{0}
\setcounter{table}{0}
\setcounter{equation}{0}
\setcounter{page}{1} % not 0 as \newpage already started a supplementary page
% References continue the numbering from the main text.

%%%%%%%%%%%%%%%% SUPPLEMENT TITLE PAGE %%%%%%%%%%%%%%%

\begin{center}
\section*{Supplementary Materials for\\ \scititle}

% Author list for the supplement
% Indicate the corresponding authors, but do NOT include institutions here
% It would be nice if the template auto-generated this, but doing so is complicated...
	Dye~SK~Sato$^{\ast}$,
	Keisuke~Yoshida,
%	Takeshi~Iinuma,
%	Masayuki~Kano,
%	Yusuke~Tanaka
\\ % we're not in a \author{} environment this time, so use \\ for a new line
\small$^\ast$Corresponding author. Email: daisukes@jamstec.go.jp\\
\end{center}

% Fill out the numbers for each type of supplementary material,
% and delete any lines that aren't applicable.
% These are just example numbers that don't match the rest of this template.
\subsubsection*{This PDF file includes:}
%Materials and Methods\\
%Supplementary Text\\
Figures S1 to S7\\%S1 and S2\
%Tables S1 to S3\\
%Captions for Movies S1 to S2\\
%Captions for Data S1 to S2
%\subsubsection*{Other Supplementary Materials for this manuscript:}
%Movies S1 to S2\\
%Data S1 to S2

\newpage

%%%%%%%%%%%%%%%% MATERIALS AND METHODS %%%%%%%%%%%%%%%

%%%%%%%%%%%%%%%% SUPPLEMENTARY TEXT %%%%%%%%%%%%%%%

%\subsection*{Supplementary Text}
%The Supplementary Text section can only be used to directly support statements made in the main text e.g. to present more detailed justifications of assumptions, investigate alternative scenarios, provide extended acknowledgements etc. Material in this section cannot claim results or conclusions that weren't mentioned in the main text. To refer to this section from the main text, just write (Supplementary Text).

%\subsubsection*{Example supplement heading}
%The two main sections of the supplement can be split up using headings.

% If your supplement is very short you might need to uncomment the following line to avoid
% layout problems with the figures and tables.
%\newpage

%%%%%%%%%%%%%%%% SUPPLEMENTARY FIGURES %%%%%%%%%%%%%%%

%\begin{figure} % Do not use \begin{figure*}
%	\centering
%    \includegraphics[width=1\textwidth]{FigS1_YamagataFukushimaSwarm_TemporalStatistics.pdf} % for an image file named example_figure.*
	% Pick an appriopriate width for the size of the image

	% Captions go below figures
%	\caption{xxxxxx}
%	\label{fig:S1} % give each figure a logical label name
%\end{figure}

\begin{figure*}[p]
\centering
 \includegraphics[width=120mm]{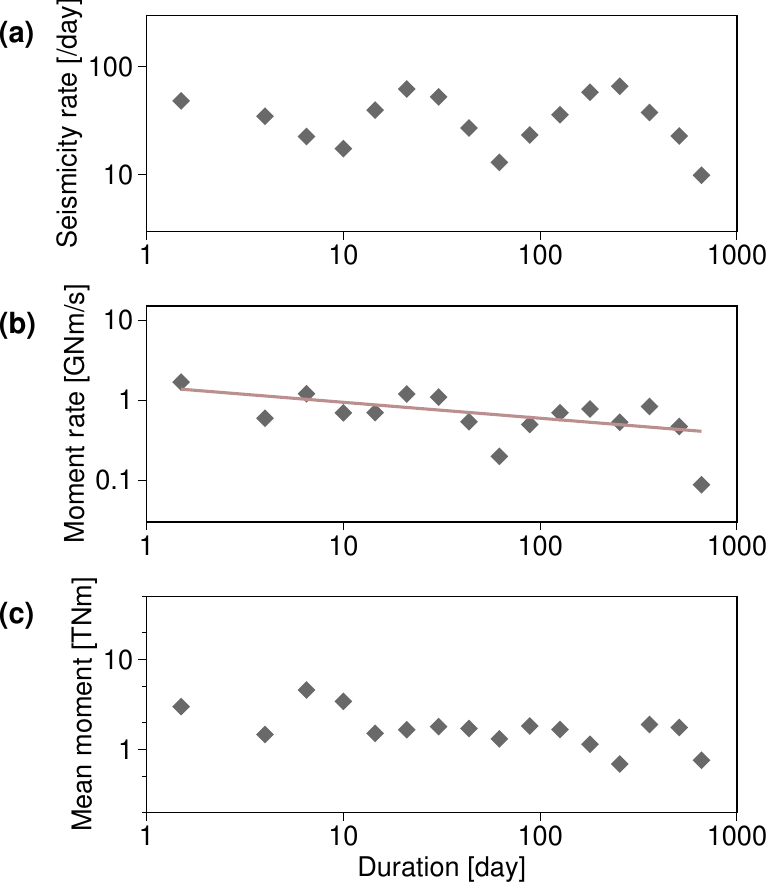}
 \caption{
Temporal evolution of swarm activity statistics: (a) seismicity rate, (b) seismic moment rate, and (c) mean seismic moment per event. The horizontal axis represents the elapsed time (duration) from the onset of the swarm activity.
All statistics are calculated within discrete log time intervals. 
The solid line indicates a power-law fit of seismic moment ($\propto t^{-0.2}$).
}
 \label{fig:S1}
\end{figure*}

\begin{figure*}[p]
\centering
 \includegraphics[width=140mm]{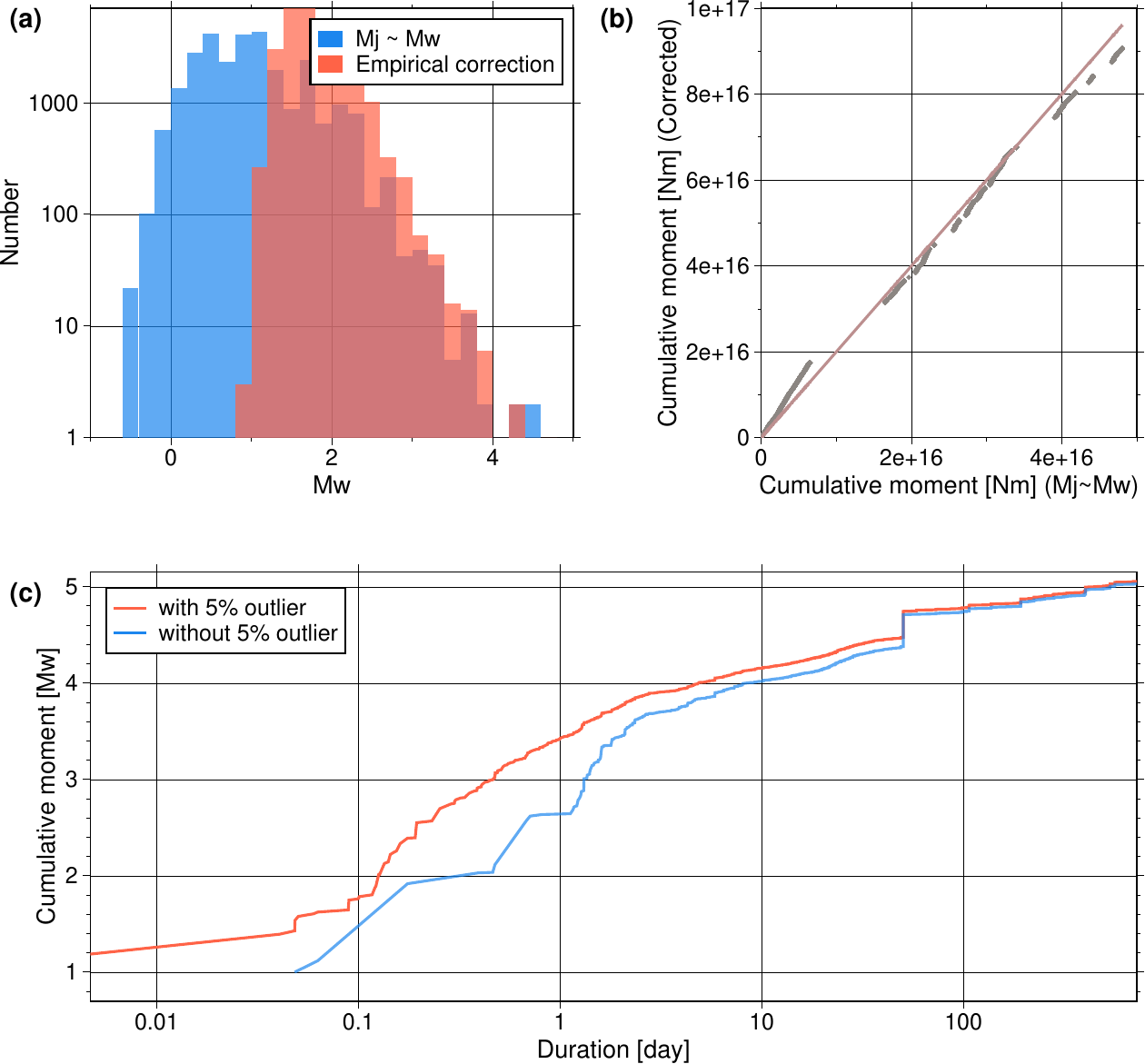}
 \caption{
Robustness checks of the seismic moment evaluation. (a) Magnitude frequency distributions. The original cataloged magnitudes assuming $M_w \approx M_{JMA}$ (blue) are compared with the corrected magnitudes following Edwards and Rietbrock (2009) (red). 
(b) Comparison of the cumulative seismic moments calculated from the original and corrected magnitudes. The solid line represents a factor-of-two difference. (c) Time evolution of the cumulative seismic moment with (red) and without (blue) the 5\% geometrical outliers excluded via Isolation Forest.
}
 \label{fig:S2}
\end{figure*}

\begin{figure*}[p]
\centering
 \includegraphics[width=150mm]{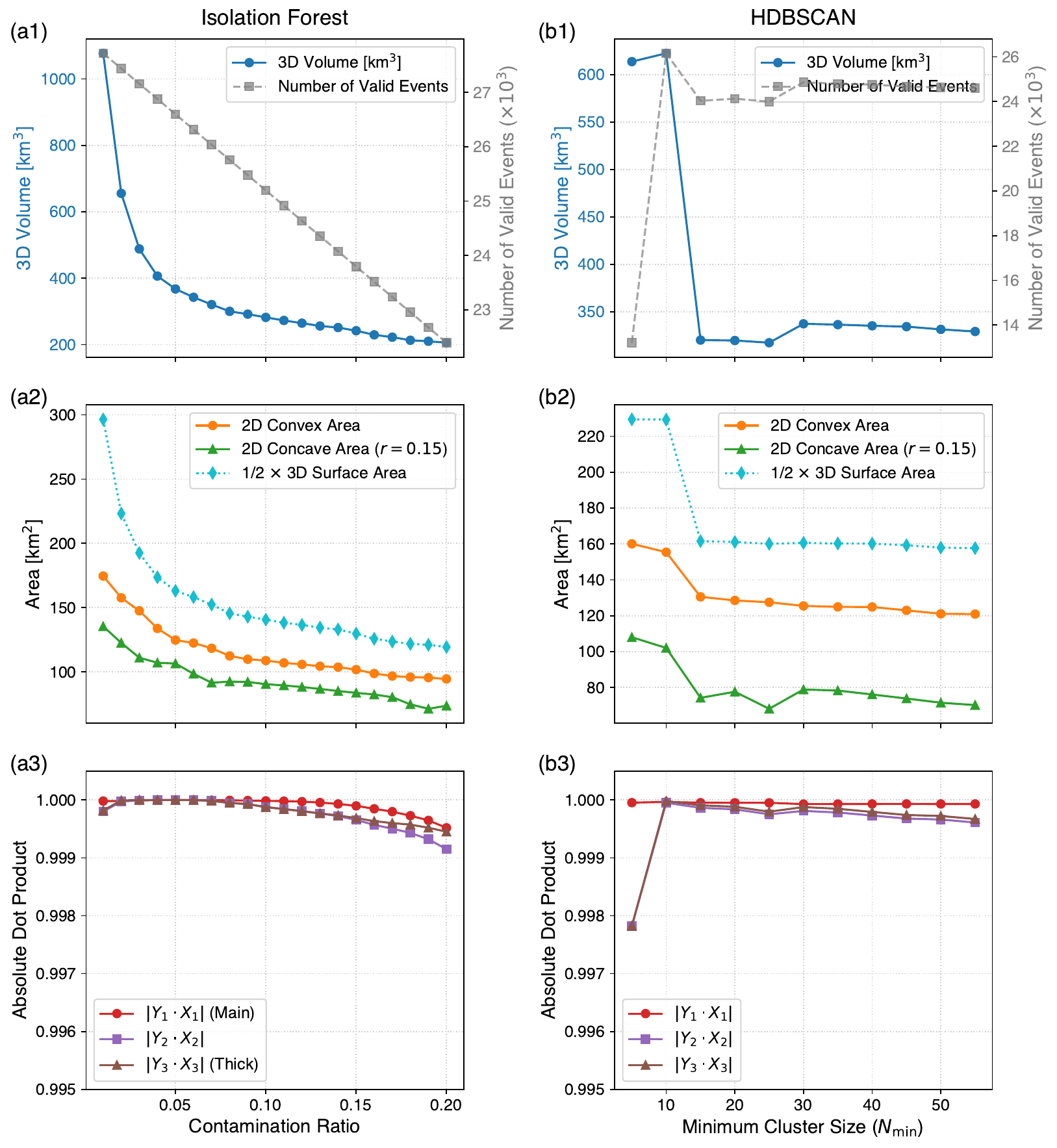}
 \caption{
Robustness check for geometrical outlier removal. The left and right columns show the dependence on the contamination ratio in the Isolation Forest and the minimum cluster size ($N_{\min}$) in HDBSCAN, respectively.
(Top row) Transition of the three-dimensional volume (blue circles) and the number of valid events (gray dashed line) derived from the filtered point cloud.
(Middle row) Transition of the projected areas: convex hull area (orange circles), concave hull area with convexity\_parameter $r=0.15$ (green triangles) on the plane spanned by the first and second principal components, and half of the surface area of the volume (cyan dotted line).
(Bottom row) Absolute inner products of the principal axes between the reference axes $X_n$ (determined at 5\% contamination in Isolation Forest) and those extracted under each parameter set $Y_n$.
}
 \label{fig:S3}
\end{figure*}

\begin{figure*}[p]
\centering
 \includegraphics[width=100mm]{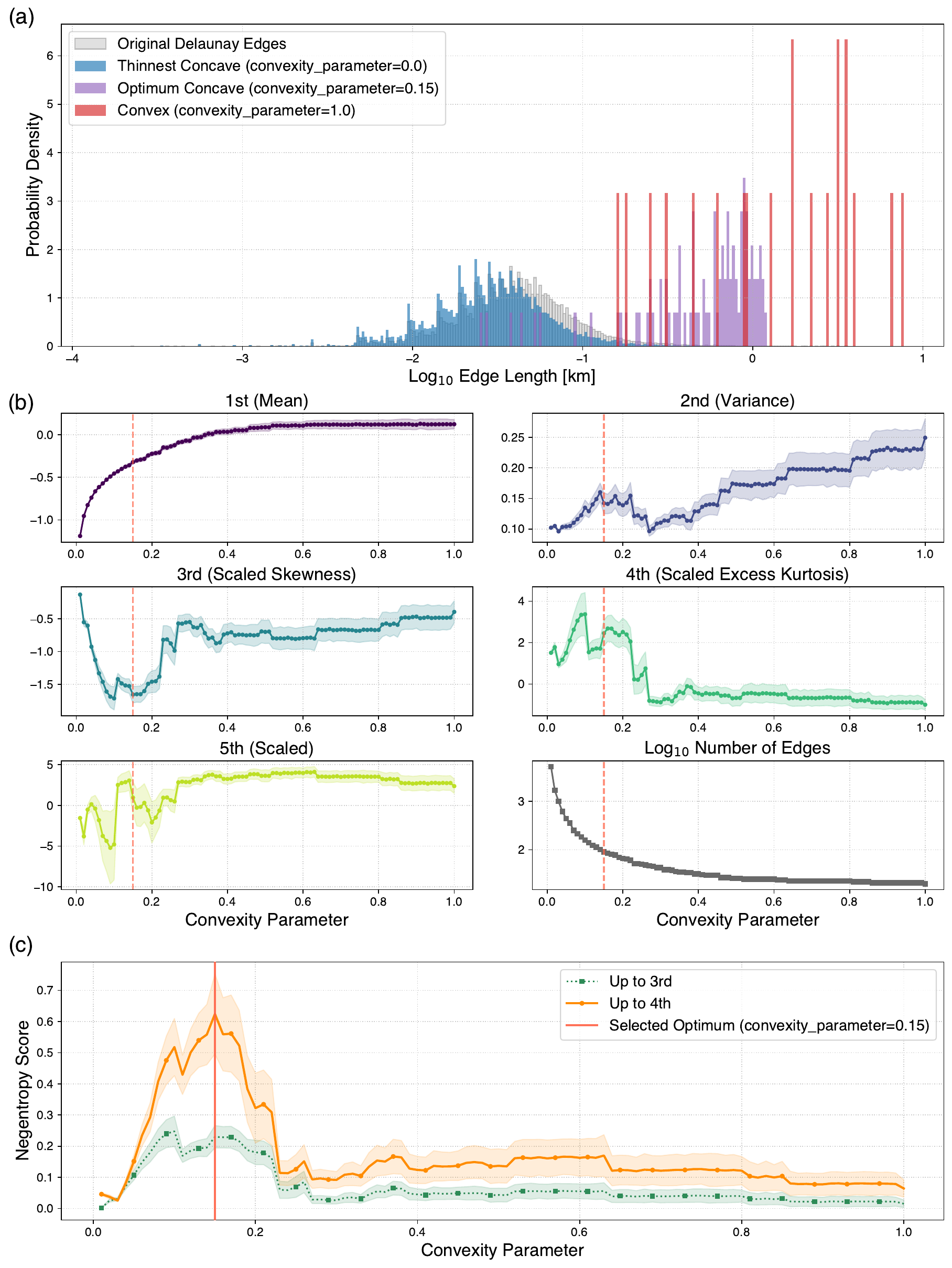}
 \caption{
Determination of the optimal concave hull via negentropy maximization.
(a) Probability density functions of $\log_{10}$ edge lengths for the original Delaunay triangulation (gray shaded) and the concave envelopes. 
The microscopic and macroscopic end-members are computed as convexity\_parameter = 0.0 (blue) and convexity\_parameter = 1.0 (red), which approximate a log-normal distribution and a mixed delta function, respectively. The maximum-negentropy distribution is also shown (convexity\_parameter = 0.15, purple). 
(b) Cumulants of the logarithmic edge length distributions and the number of edges as functions of convexity\_parameter. Standardized cumulants ($\kappa_i$) are shown for the third-to-fifth orders.
Solid lines and shaded areas denote the ensemble means and $\pm 1\sigma$ standard errors evaluated via subsampling (500 iterations of 80\% non-replacement extraction). The optimal convexity\_parameter is indicated by vertical lines.
(c) Negentropy $J$ (distance from the Gaussian distribution with the same mean and variance), approximated via Edgeworth expansions. The expansion series is evaluated for the first and second orders ($J \simeq \kappa_3^2/12$ and $J \simeq \kappa_3^2/12 + \kappa_4^2/48+
7\kappa_3^4/48 -\kappa_3^2\kappa_4/8$, respectively). The global maximum of the second order expansion is indicated by the vertical line. 
}
 \label{fig:S4}
\end{figure*}

\begin{figure*}[p]
\centering
 \includegraphics[width=150mm]{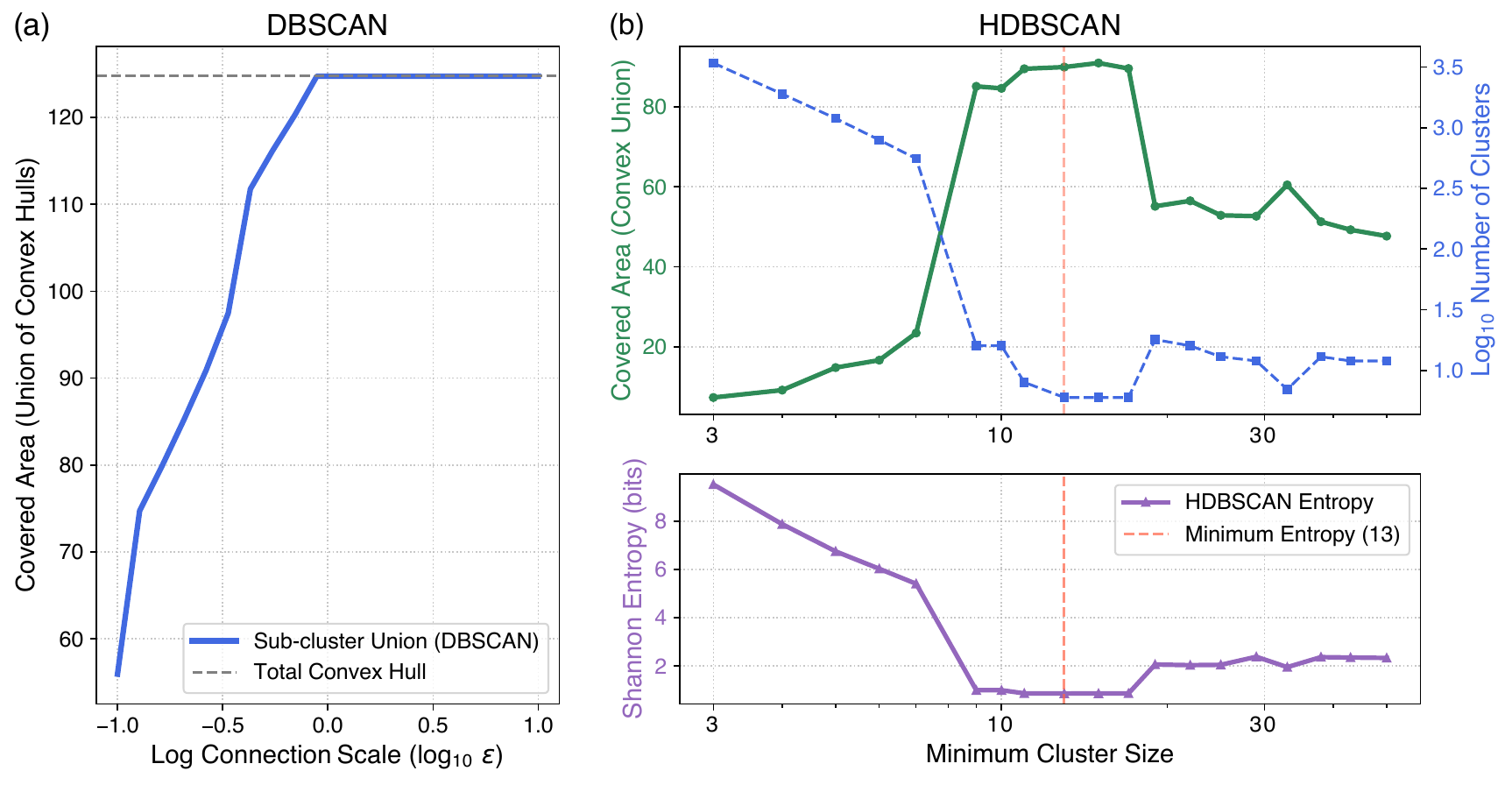}
 \caption{
Event clustering via entropy minimization.
(a) Covered areas (the union of sub-cluster convex hulls) of DBSCAN clusters as functions of the connection scale ($\epsilon$). Due to the long-tailed nature of the log-normal distribution of inter-event distances, the covered area gradually increases with $\epsilon$ without forming a robust plateau, making it difficult to stably detect the intermediate regime. 
(b) HDBSCAN clustering metrics as functions of the minimum cluster size ($N_{\min}$). The top panel shows the covered area (solid green) and the number of valid clusters (dashed blue). The bottom panel shows the Shannon partition entropy (solid purple). The minimum entropy state ($N_{\min}=13$, dashed vertical) is located at the plateau region of all metrics.
}
 \label{fig:S5}
\end{figure*}

\begin{figure*}[p]
\centering
 \includegraphics[width=150mm]{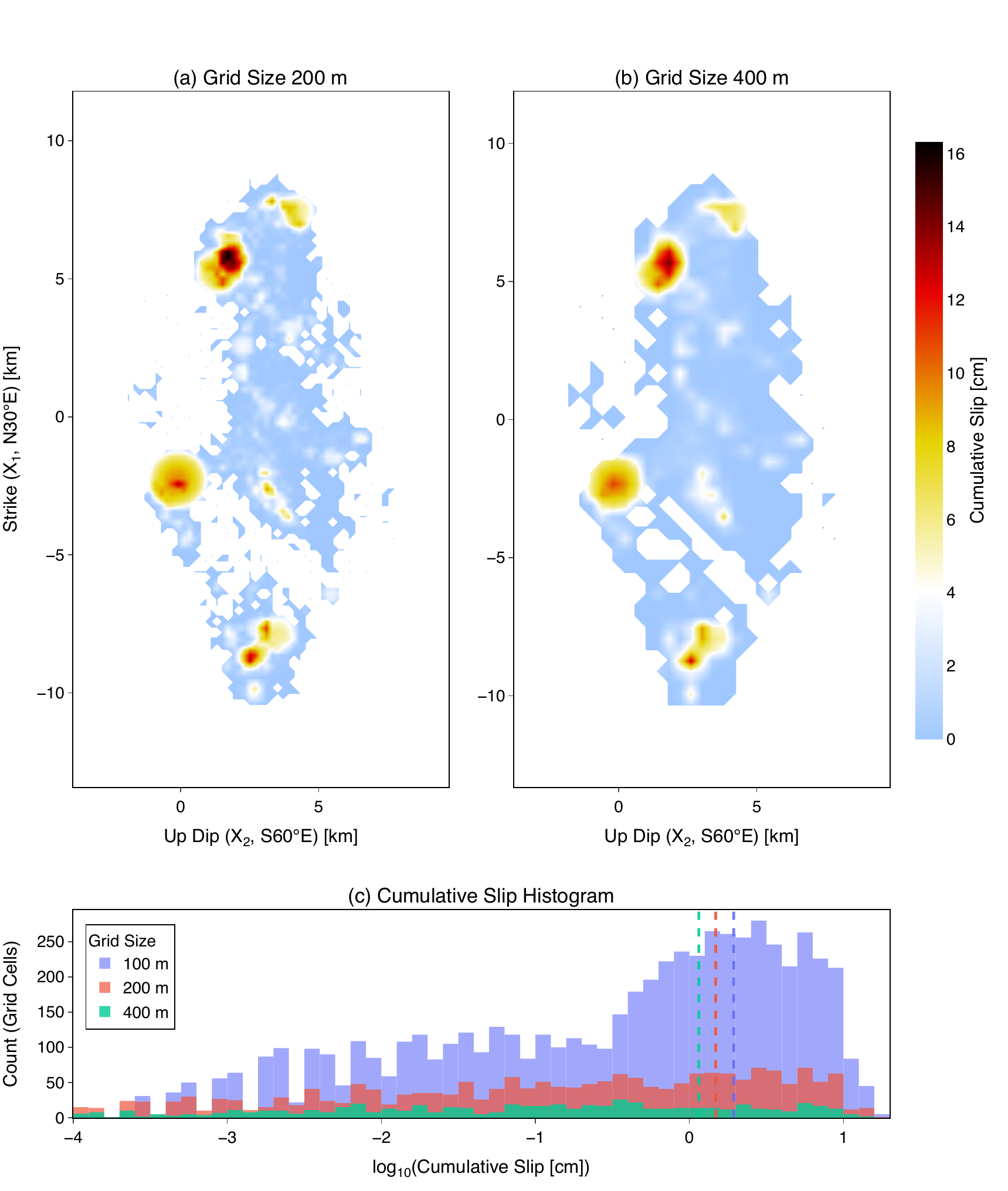}
 \caption{
Grid-size dependence of slip distributions based on the circular crack model. The modeled fault plane consists of $X_1$ (N30${}^\circ$E, strike) and $X_2$ (S60${}^\circ$E, up-dip 15${}^\circ$) axes. The spatial distribution of cumulative slip distance is computed for two different grid resolutions (a: 200 m; b: 400 m). Histograms of the logarithmic cumulative slip distance are also shown (c; values below 1 µm are excluded from the histogram plot). The vertical lines, colored correspondingly in the histograms, indicate the average cumulative slip distances of nonzero-slip grid cells.
}
 \label{fig:S6}
\end{figure*}

\begin{figure*}[p]
\centering
 \includegraphics[width=150mm]{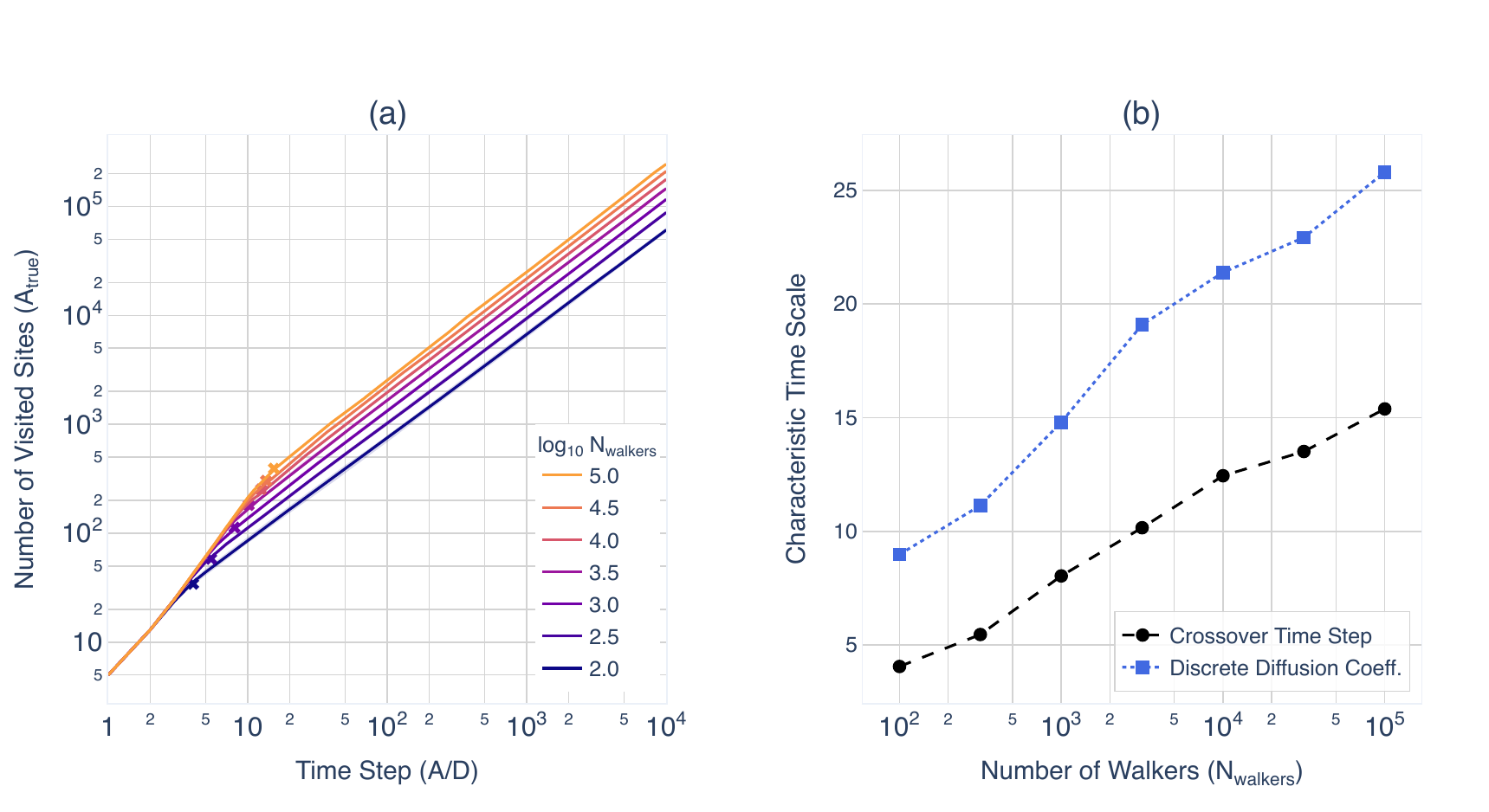}
 \caption{
Brownian rupture model for different numbers of random walkers.
(a) Time evolution of the cumulative number of visited sites, representing the discrete true rupture area ($A_{\rm true}$), for different numbers of random walkers ($N_{\rm walkers}$). Solid lines and shaded regions represent the ensemble averages and standard deviations over $N_{\rm ens} \approx 600/\sqrt{N_{\rm walkers}}$ realizations, respectively. Crosses indicate the crossover points determined by the intersection of the linear fits in the log-log space for the early ballistic ($A_{\rm true} \propto t^2$) and late diffusive ($A_{\rm true} \propto t$) regimes. (b) Dependence of the discrete crossover time step $t_*$ (black dashed line) and the macroscopic diffusion coefficient $D$ (blue dotted line) on $N_{\rm walkers}$.
}
 \label{fig:S7}
\end{figure*}

%%%%%%%%%%%%%%%% SUPPLEMENTARY TABLES %%%%%%%%%%%%%%%

%\begin{table} % Do not use \begin{table*}
%	\centering
%	% Captions go above tables
%	\caption{\textbf{Estimates of exp-plus-log fits.}
%		Least-square estimates of $t_2$ and $t_{\rm c}$ are listed for the case using an empirical function $F_2$ (eq.~2).  }
%	\label{tab:sup3} % give each table a logical label name
%
%	\begin{tabular}{c|ccc} % four columns, alignment for each
%		%\\
%		\hline
%		$F_2$ fit (eq.~2) & Post & Co\&Post & Total \\
%		\hline
%		$t_2$ [date] & 2010-01-27$\pm$1 & 2010-02-01$\pm$1 & 2010-01-23$\pm$5 \\
%		$t_{\rm c}$ [days] & 87$\pm$5 & 46$\pm$2 & 593$\pm$152\\
%		$t_2^\prime$ [date] & 2010-02-26 & 2010-03-11 & 2010-05-27\\
%		%\hline
%	\end{tabular}
%\end{table}

\end{document}